# Observation of optomechanical coupling in a microbottle resonator

*Motoki Asano,[1] Yuki Takeuchi,[1] Weijian Chen,[2] Şahin Kaya Özdemir,[1,2]\* Rikizo Ikuta,[1] Nobuyuki Imoto,[1] Lan Yang,[2] and Takashi Yamamoto[1] \**

\*Corresponding Author: ozdemir@wustl.edu; yamamoto@mp.es.osaka-u.ac.jp

[1]Graduate School of Engineering Science, Osaka University, Toyonaka, Osaka, 560-8531, Japan
[2]Department of Electrical and Systems Engineering, Washington University in St. Louis, MO 63130, USA

In this work, we report optomechanical coupling, resolved sidebands and phonon lasing in a solid-core microbottle resonator fabricated on a single mode optical fiber. Mechanical modes with quality factors ($Q_m$) as high as $1.57 \times 10^4$ and $1.45 \times 10^4$ were observed, respectively, at the mechanical frequencies $f_m = 33.7 \text{ MHz}$ and $f_m = 58.9 \text{ MHz}$. The maximum $f_m \times Q_m \sim 0.85 \times 10^{12}$ Hz is close to the theoretical lower bound of $6 \times 10^{12}$ Hz needed to overcome thermal decoherence for resolved-sideband cooling of mechanical motion at room temperature, suggesting microbottle resonators as a possible platform for this endeavor. In addition to optomechanical effects, scatter-induced mode splitting and ringing phenomena, which are typical for high-quality optical resonances, were also observed in a microbottle resonator.

1. Introduction

Interaction between optical and mechanical modes via radiation pressure of photons has been of increasing interest not only for fundamental understanding of momentum transfer from photons to massive objects but also for a variety of applications ranging from precision measurement of small forces, mass and displacement, gravitational wave detection and controlling light propagation via phonons to creating nonclassical states of light and mechanical motion, interfacing photonic and solid-state qubits and building hybrid quantum systems [1]. With the availability of high finesse optical micro- and nano-resonators supporting mechanical modes, the field of optomechanics has advanced rapidly reaching the levels where coherent interactions between optical and mechanical modes are controlled to experimentally study laser cooling [2-4], optomechanically induced transparency [5-7], Rabi oscillation [8], Ramsey interference [9], coherent wavelength conversion [10,11], phonon lasing [12,13] and classical and quantum correlation measurement using read/write operatoin [14,15]. Similarly, experiments have been performed to study nonlinear dynamics and chaos [16-20] and optofluidic sensing [21,22] in optomechanical systems. Optomechanics has been studied theoretically for its possible applications and uses in quantum information processing [23-25] and gravitational wave detection [26], as well as for the realization of PT-symmetric systems [27,28].

The fundamental mechanism that leads to coupling between the mechanical motion and the intracavity optical field in a WGM resonator is the direct momentum transfer from intracavity photons (i.e., radiation pressure) to the vibrational mechanical modes of the resonator structure as the photons propagate along the boundary of the resonator via total internal reflection. This, in turn, leads to distortions of the structure and modifies the optical path length that shifts the optical resonance frequency. In other words, the back-action of the mechanical motion creates sidebands on the intracavity field that are separated by the multiples of the mechanical frequency, $f_{\text{cav}} \mp nf_{\text{m}}$ where $n = 0,1,2,...$ and $f_{\text{cav}}$ is the resonance frequency of the cavity. The first pair of sidebands located at $f_{\text{cav}} - f_{\text{m}}$ and $f_{\text{cav}} + f_{\text{m}}$ are called as the Stokes and anti-Stokes, respectively. **Figure 1a** depicts an illustration of a bottle resonator with the pump, Stokes and anti-Stokes modes. If the frequency $f_{\text{laser}}$ of the pump laser is blue (red)-detuned from the cavity resonance frequency $f_{\text{cav}}$ by $f_{\text{m}}$, mechanical motion is amplified (suppressed) and heated (cooled), and Stokes (anti-Stokes) is enhanced.

Among many physical systems, such as nanomechanical resonators [29], cavities with a membrane in the middle [30], Fabry-Pérot cavities [31], photonic crystal cavities [3], whispering-gallery-mode (WGM) resonators (e.g., microspheres [32], microdisks [33], microtoroids [34] and microdroplet [35]) have emerged as versatile platforms for optomechanics due to the presence of optical and mechanical modes in a microscale device with high quality factors [36,37]. In the past few years, WGM microresonators with a bottle-like geometry (Figure 1a) have attracted interest due to their ultra-high optical quality factors $Q_{\text{opt}} \geq 10^8$, 3D confinement of optical modes and ease of fabrication [38,39]. Recent experiments of cavity-QED with solid-core bottle resonators fabricated from single mode optical fibers [40], and optomechanics using hollow-core bottle resonators fabricated from capillaries [41] have shown that bottle resonators are very promising for applications and fundamental studies.

In this paper, we report the observation of optomechanical coupling, phonon lasing and resolved sidebands in solid-core bottle resonators. Our results show that solid-core microbottle resonators are very promising for achieving resolved-sideband cooling at room temperatures.

2. Experiments and results

The bottle resonators used in our experiments were fabricated from standard single mode optical fibers (10 μm core and 125 μm cladding) by heat-and-pull method. We heated the optical fiber using an oxygen-hydrogen flame and tapered it at two separated points which formed the necks of the bottle. In this way, we prepared three microbottle resonators (referred to as µB1, µB2, and µB3) whose dimensions were roughly estimated from their optical microscope images, respectively, as (100 µm × 30 µm × 4 mm) and (105 µm × 48 µm × 4 mm), and (77 µm × 26 µm × 4 mm) where the first, second and third entries in the parentheses corresponding to the maximum diameter of the body, the diameter of the necks, and the distance between the two necks of each bottle resonator. An illustration of the experimental setup we used to characterize the optomechanical coupling in the fabricated microbottle resonators is given in Figure 1b.

Light from a tunable external cavity diode laser (ECDL) having a linewidth of 300 kHz in the 1550 nm band was amplified by an erbium-doped fiber amplifier (EDFA) and was used as the pump to excite the mechanical oscillations. The polarization and the power of the pump light were controlled by a fiber-based polarization controller (PC) and an attenuator (Att.), respectively. The pump light was guided into the WGMs of the bottle resonators using a tapered fiber which was also fabricated using the heat-and-pull method. The coupling between the tapered fiber and the bottle resonator was controlled by changing the distance between them using a 3D nanopositioning system. The intracavity light field was coupled out to the same tapered fiber and transmitted to a photodetector (PD).

2.1 Mode-splitting and ringing in microbottle resonators

We first characterized the optical modes of the resonators by obtaining their transmission spectra by scanning the wavelength of the ECDL while monitoring the output of the PD with an oscilloscope. For this purpose the power of the light input to the resonators was set to very low levels such that no mechanical oscillation or thermal effect was observed. **Figure 2** depicts typical transmission spectra for µB1 (Fig. 2a), µB2 (Fig. 2b) and µB3 (Fig. 2c), obtained when the laser frequency was scanned linearly. The overshoot peaks over the unity transmission are caused by the power buildup in the cavity, which is often observed in the ultra-high Q regime. The insets show the close-up spectra of the resonances seen in the transmission spectra. Scatter-induced mode splitting [42-47] and the ringing phenomenon [48], which are typical for ultra-high-Q resonances but have not been reported for bottle resonators up to now, are clearly seen. The observed mode-splitting was about 4.2 MHz and the quality factors of the split modes were about $1.3 \times 10^8$ and $1.4 \times 10^8$ for µB1. The resonance depicting the ringing phenomenon had a quality factor of $3.2 \times 10^8$, estimated using the method given in Ref [48]. The other resonance mode colored in blue in Figure 2a had a quality factor of $1.3 \times 10^8$. For µB2, on the other hand, the mode-splitting was about 2.0 MHz and the quality factors of the split modes were about $1.4 \times 10^8$ and $1.9 \times 10^8$ (Fig. 2b). The resonance depicting the ringing phenomenon had a quality factor of $4.1 \times 10^8$, and the other resonance mode colored in blue in Figure 2b had a quality factor of $2.0 \times 10^8$. For µB3, the mode-splitting was about 5.1 MHz and the quality factors of the split modes were about $5.4 \times 10^7$ and $9.8 \times 10^7$ (Fig. 2c). The other resonance mode colored in blue in Figure 2c had a quality factor of $3.7 \times 10^7$. There was no ringing phenomenon in µB3 because the observed resonance modes had lower quality factors.

2.2 Optomechanical oscillations in microbottle resonators

Next at elevated input powers we characterized the optomechanical coupling by monitoring the output of the PD with an electric spectrum analyzer (ESA). During the measurements for optomechanical coupling, the tapered fiber was kept in contact with the bottle resonator. This helped to minimize, if not eliminate, the fluctuations of the coupling condition at the expense of increased mechanical mass which affected the power threshold to observe mechanical oscillations. The spectrum obtained at the ESA corresponds to the spectrum of the beat note signal obtained by the mixing of the pump light and the scattered light caused by the excitation of the mechanical oscillation. In Figure 1b we present a typical spectrum obtained at the ESA showing a mechanical mode with a frequency of $f_m = 35.6$ MHz.

Our numerical simulations with 3D finite element method (FEM) in COMSOL Multiphysics for the bottle resonator used in our experiments (dimensions are provided above) revealed seven fundamental mechanical modes around the mechanical frequency of $f_m = 35.6$ MHz, having different frequencies and spatial displacement patterns (**Figure 3**). Together with high order mechanical modes the total number of modes observed in the simulations within the frequency range of 7-45 MHz was 21 (see figure and video in online supporting information). The mechanical mode at 36.4 MHz is close to the mechanical frequency of 35.6 MHz observed for the resonator used in the experiments. As seen in Figure 3 the bottle resonator experiences deformation with the radial vibration around its center, corresponding to a radial breathing mode (RBM), which has been reported previously for microtoroids [49] and microspheres [32]. In the following, we focus on this RBM and investigate its characteristics.

**Figure 4** show the spectra of the mechanical oscillations at 35.6 MHz in µB1, 33.7 MHz in µB2, and 58.9 MHz in µB3, together with the higher harmonics at various values of the optical pump power. The first sharp peak in each of the spectra corresponds to the fundamental mode of the mechanical oscillation, and the peaks appearing with an interval of $f_m$ from the fundamental mode are the harmonics ($f_m = 35.6$ MHz, $f_m = 33.7$ MHz, and $f_m = 58.9$ MHz for µB1, µB2, and µB3, respectively). At an optical pump power of 24.1 mW, we observed up to the thirteenth harmonics which is located at 462 MHz for µB1 (Figure 4Ia).

When the optical pump power was decreased, the higher harmonics of the mechanical oscillation gradually disappeared, i.e., at a power of 20.1 mW up to only the fifth harmonic was observed (Figure 4Ib-d). For the fundamental mode at 33.7 MHz of µB2, we observed up to the fifteenth harmonics which is located at 505 MHz at a pump power of 21.2 mW, and up to the seventh harmonics when the pump was decreased to 8.7 mW (Figures 4IIa-d). For the mechanical oscillation in µB3 whose fundamental mode is located at 58.9 MHz (Figure 4III), we observed up to the seventh harmonics which is located at 412 MHz (Figure 4IIIa) at a pump power of 31.4 mW. When the power was decreased to 16.8 mW, we observed up to the third harmonics (Figure 4IIId). These results imply that mechanical modes with high oscillation thresholds require more pump power to observe higher order harmonics.

2.3 Phonon lasing in microbottle resonators

Plot of the peak power spectrum density (PSD) versus optical pump power reveals laser threshold behavior not only for the fundamental mode but also for the harmonics of the 35.6 MHz mode in µB1 (**Figure 5Ia** and the insets), of the 33.7 MHz mode in µB2 (**Figure 5IIa** and the insets), and of the 58.9 MHz mode in µB3 (**Figure 5IIIa** and the insets). The oscillation threshold for the fundamental mode of 35.6 MHz mode in µB1 was obtained as $P_1 = 13.3$ mW by extrapolation in the measured pump power vs peak PSD data. The threshold pump powers for the second, third and fourth harmonics were $P_2 = 13.4$ mW, $P_3 = 14.1$ mW, and $P_4 = 14.2$ mW, respectively (Figure 5Ia). For the fundamental mode of 33.7 MHz mode in µB2, we obtained the lasing thresholds as $P_1 = 7.4$ mW, $P_2 = 8.1$ mW, $P_3 = 8.2$ mW, and $P_4 = 8.4$ mW, respectively, for the fundamental, second, third and fourth harmonics (Figure 5IIa). For the fundamental mode of 58.9 MHz mode in µB3, we obtained the lasing thresholds as $P_1 = 10.7$ mW, $P_2 = 10.9$ mW, $P_3 = 11.0$ mW, and $P_4 = 11.2$ mW, respectively, for the fundamental, second, third and fourth harmonics (Figure 5IIIa). These results clearly show that the threshold optical pump power for higher order mechanical oscillation increases with the order. In addition to the threshold behavior, we observed gain saturation with the pump power of $P_{saturation} = 9.0$ mW for the fundamental mode in µB2.

In order to compare the threshold powers measured in our experiments with theoretical expectations, we used a model developed by Kippenberg *et al* [34] which relates threshold power for phonon lasing with the experimentally accessible parameters and the mechanical mass of the resonator as

$$P_{th} = \frac{8\pi^2 f_m^2 f_{cav} R^2 m_{eff}}{Q_m Q_0 Q_L} \frac{\left|1 + \left(\frac{Q_0}{Q_L} - 1\right) + \frac{2i\delta Q_0}{f_{cav}}\right|}{4\left(\frac{Q_0}{Q_L} - 1\right)}$$

$$\times \left[\frac{1}{1+4\left(\frac{Q_L}{f_{cav}}\right)^2(\delta+f_m)^2} - \frac{1}{1+4\left(\frac{Q_L}{f_{cav}}\right)^2(\delta-f_m)^2}\right]^{-1}. \quad (1)$$

where $f_m$ is the frequency of the mechanical mode, $f_{cav}$ is the frequency of the resonance mode of the WGM, $R$ is the radius of the microresonator, $m_{eff}$ is the effective mechanical mass of the resonator, $\delta$ is the frequency detuning between the pump laser and the cavity resonance, and $Q_m$, $Q_0$, and $Q_L$ respectively corresponds to the quality factors of the mechanical mode, and the intrinsic and the loaded quality factors of the optical mode of the microresonator. Here, $Q_0$ takes into account all the losses (e.g., scattering, material absorption, radiation losses, etc.) except the coupling losses, and $Q_L$ takes into account both $Q_0$ and the coupling quality factor.

Expression in Eq. (1) suggests that using the experimentally accessible values $R, f_{cav}, f_m, Q_0, Q_L, Q_m$ and $\delta$, one can estimate $m_{eff}$ from the measured threshold power of phonon lasing. Similarly, if $m_{eff}$ is known, one can estimate the phonon-lasing threshold power to compare with the experimentally obtained value.

In our experiments, we do not have direct access to $\delta$ because we use the thermal locking method which prevents us to know the exact value of $\delta$. However, we can estimate $m_{\text{eff}}$ by numerical simulation in COMSOL according to the procedure given in Refs. [50,51] using the dimensions and material properties of the fabricated microbottles. For our microbottle resonators, we estimated $m_{\text{eff}}$ as $2.9 \times 10^{-9}$ kg, $3.5 \times 10^{-9}$ kg and $1.1 \times 10^{-9}$ kg, respectively, for µB1, µB2 and µB3. Using these values in Eq. (1) together with the experimentally accessible parameters, we determined the relation between the phonon lasing threshold $P_{\text{th}}$ and the detuning $\delta$ for each of the resonators and plotted in **Fig. 6.** It is clearly seen that the lowest lasing threshold is obtained when the detuning $\delta$ is equal to the frequency of the mechanical mode $f_{\text{m}}$. At the optimal detuning point ($\delta = f_{\text{m}}$), we estimate the phonon lasing thresholds of µB1, µB2 and µB3 as 0.64 mW, 0.77 mW, and 0.86 mW, respectively. These threshold powers are more than twenty-fold, ten-fold and twelve-fold smaller than the measured threshold powers of 13.1 mW, 7.4 mW and 10.7 mW for µB1, µB2 and µB3. Inserting the threshold powers obtained in our experiments onto the threshold versus detuning curves given in Fig. 6, we find that our experiments were performed at non-optimal detuning conditions with two possible detuning values, one of which is on the blue and the other on the red side of the optimal detuning (i.e., $\delta = f_{\text{m}}$). Since we performed the experiments at the detuning where the thermal locking was observed for the first time while the wavelength of the pump laser was scanned from shorter to longer wavelength, the detuning present in our experiments should be on the blue side of the optimal detuning (i.e. far-detuned from the cavity resonance). Thus, we estimate the effective detuning in our experiments as $\delta = 50$ MHz (filled circles in Fig. 6), $\delta = 37$ MHz (filled square. in Fig. 6), and $\delta = 72$ MHz (filled star in Fig. 6), respectively for the microbottle resonators µB1, µB2 and µB3. It is clear that performing the experiments at optimal detuning or close to the optimal detuning will significantly decrease the threshold power for phonon lasing in the microbottle resonators.

Another indication of the laser-like behavior is the narrowing of linewidth above the threshold when compared to the linewidth below the threshold. The spectra of the mechanical oscillation at 35.6 MHz in µB1, 33.7 MHz in µB2 and 58.9 MHz in µB3 obtained below and above the lasing thresholds of the fundamental modes are given in Figures 5Ib, 5IIb, 5IIIb and 5Ic, 5IIc, 5IIIc, respectively, which clearly show the narrowing of the linewidths above lasing threshold. Note that the linewidths of the oscillations above the lasing threshold shown in Figure 5Ic, 5IIc and 5IIIc were so narrow that it could not be resolved with our ESA.

2.4 Estimating intrinsic quality factors of mechanical modes and frequency-Q products

The net loss $\gamma$ of a mechanical mode is determined by the interplay between two components. First is the intrinsic mechanical loss $\gamma_{int}$ which reflects the properties of the material, used to fabricate the device, and the surrounding where the device is operated or measured. Second is the optomechanical gain $\gamma_{opt}$ induced via the coupling between the optical and mechanical modes. This can be probed using a blue-detuned pump, that is when the system is heated (mechanical amplification). The spectra shown in Fig.5 correspond to phonon lasing (i.e., amplification of mechanical oscillations) where optomechanical gain surpasses the mechanical losses in the system.

The intrinsic quality factor of the mechanical mode can be estimated by measurements performed well-below the lasing threshold. Because the optomechanical coupling is proportional to the intracavity photon number [1], the intrinsic mechanical loss can be estimated by extrapolating from the experimentally obtained dependence of the linewidth of the mechanical mode on the pump power. In order to do this, we used a stable distributed feedback (DFB) laser (central wavelength of 1542 nm and linewidth of 2.5 kHz) as a pump to induce mechanical oscillations. The pump powers were chosen to be smaller than the lasing threshold. We used only µB2 and µB3 in these experiments, because µB1 became contaminated, and the optical and mechanical resonances, for which the results were depicted in Figures 2, 4 and 5, were lost. **Figure 7Ia** and Figure 7IIa show typical PSD spectra obtained for the mechanical oscillation at $f_m$ = 33.7 MHz in µB2 and $f_m$ = 58.9 MHz in µB3, respectively, at three different pump powers. It is clearly seen as the pump power decreases, the linewidths become larger and peak PSD decrease. Figure 7Ib and Figure 7IIb show that there is an inverse relation between the linewidths of the mechanical modes and the pump powers, that is linewidths increase linearly with decreasing pump power. These results agree well with the prediction of the theoretical model [1]. By extrapolating the experimental data shown in Figure 7Ib and Figure 7IIb to the zero pump power, we find the intrinsic linewidths of the mechanical modes of µB2 and µB3 as 2.15 kHz and 4.05 kHz, respectively, which correspond to the mechanical quality factors of $1.57 \times 10^4$ and $1.45 \times 10^4$.

Theoretical considerations imply that in order to overcome the thermal decoherence in one cycle of the mechanical oscillation at room temperature, the frequency-Q product of the mechanical mode should be greater than $6 \times 10^{12}$ Hz [1]. Previously frequency-Q product greater than $10^{12}$ Hz was realized in various optomechanical and electromechanical systems, such as $1.8 \times 10^{12}$ Hz in a spoke-anchored WGM toroidal resonator [7], $2 \times 10^{13}$ Hz in a cavity with a $Si_3N_4$ membrane in the middle [52], $3.9 \times 10^{14}$ Hz in a patterned Si nanobeam with external phononic bandgap shield [53], and $3.5 \times 10^{12}$ Hz in an electromechanical system with a thin aluminum membrane parametrically coupled to a superconducting microwave resonant circuit [54]. The frequency-Q products of the mechanical modes in our microbottle resonators μB2 and μB3 are calculated as $33.7 \text{ MHz} \times 1.57 \times 10^4 = 0.53 \times 10^{12}$ Hz and $58.9 \text{ MHz} \times 1.45 \times 10^4 = 0.85 \times 10^{12}$ Hz, respectively. These values surpass the value of $0.46 \times 10^{12}$ Hz evaluated at the sub-threshold pump power for a hollow-core microbottle resonator [41], and are close to the theoretical lower limit of $6 \times 10^{12}$ Hz for quantum cavity optomechanics.

2.5 Resolving sidebands in the transmission spectra

Resolved-sideband cooling that holds the potential for cooling phonons into the ground state can be experimentally observed if the linewidth $\kappa/2\pi$ of the optical resonance is smaller than the mechanical oscillation frequency $f_m$. In order to achieve resolved-sideband cooling with bottle resonators, we should first be able to resolve sidebands in bottle resonators. We tested our resonators for resolved sideband using a pump and probe scheme in which the DFB laser (central wavelength of 1542 nm and linewidth of 2.5 kHz) was used to pump the mechanical vibrations and an ECDL (1550 nm band and linewidth of 300 kHz) whose wavelength was scanned linearly was used for probing the sidebands. The pump and probe lights were coupled into and out of the resonator via the same fiber taper. The light out-coupled and transmitted through the fiber was passed through a tunable optical band-pass filter to separate the pump and the probe lights. We have identified optical resonances with $\kappa/2\pi = 5.4$ MHz, $\kappa/2\pi = 4.6$ MHz and $\kappa/2\pi = 10.4$ MHz in μB1, μB2 and μB3. With the respective mechanical oscillation frequencies of 35.6 MHz, 33.7 MHz and 58.9 MHz, we find that $\kappa/2\pi < f_m$ is satisfied in all the microbottle resonators for resolving sidebands. The experimentally-obtained spectra given in (**Figure 8**) for the probe fields in each of the microbottle resonators clearly showed the resolved sidebands. With increasing pump powers, the enhanced mechanical oscillations not only lead to deeper sidebands but also help to observe higher-order sidebands as shown in **Figure 9** for μB1.

3. Conclusion and Outlook

In conclusion, we report the observation of several interesting features for solid-core microbottle resonators. These include scatter-induced mode splitting and ringing phenomena in ultra-high quality factor optical modes, coupling between the optical and mechanical modes supported in the same resonator, phonon lasing and the resolved sideband spectrum. These features imply the possibility of high-performance optical sensing, highly efficient optomechanical heating (amplification) and cooling, and more importantly cavity optomechanics in the quantum regime using microbottle resonators. These findings open new venues for microbottle resonators for applications and for fundamental studies.

In particular, downsizing the dimensions of microresonator led to the improvement of frequency-Q product. With this strategy resonator supporting mechanical modes with higher frequency and quality factor can be fabricated. In addition, introducing microbottle resonators used in our experiments into vacuum environment is expected to improve the frequency-Q product by increasing intrinsic mechanical quality factor. These imply that optimizing the structure and the environment of microbottle resonators will help to overcome the theoretical limitation for quantum cavity optomechanics.

**Acknowledgements**
This work was supported by MEXT/JSPS KAKENHI Grant Number 16H01054, 16H02214, 15H03704, 15KK0164, and Program for Leading Graduate Schools: "Interactive Materials Science Cadet Program". SKO, LY and WC are supported by the Army Research Office under grant No. W911NF-16-1-0339.

**FIGURE 1**

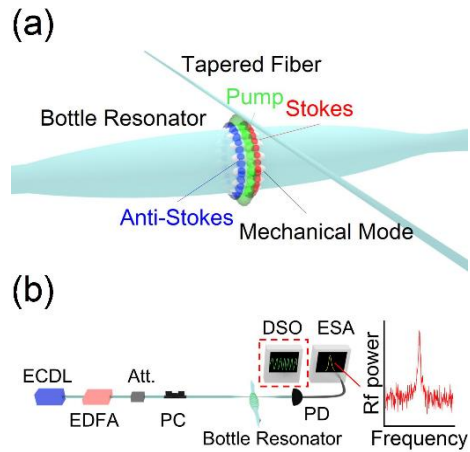

**Figure 1.** (a) An illustration of a solid-core microbottle whispering-gallery-mode (WGM) resonator for optomechanics. Optomechanical coupling between the optical pump mode and the mechanical mode creates the Stokes and Anti-Stokes fields which are, respectively, red- and blue-detuned from the pump field. (b) A schematic illustration of the experimental setup. Light from an external cavity diode laser (ECDL) was amplified by an erbium-doped fiber amplifier (EDFA) and then coupled into and out of the WGMs via a tapered fiber. Transmission spectrum was monitored by a photodiode (PD) connected to a digital storage oscilloscope (DSO) and an electrical spectrum analyzer (ESA). An attenuator (Att.) and a polarization controller (PC) were used to adjust the power and the polarization of the pump field coupled into the microbottle resonator, respectively. The radio frequency (Rf) power versus frequency spectrum in red is a typical spectrum acquired by the ESA denoting a mechanical oscillation at 35 MHz.

**FIGURE 2**

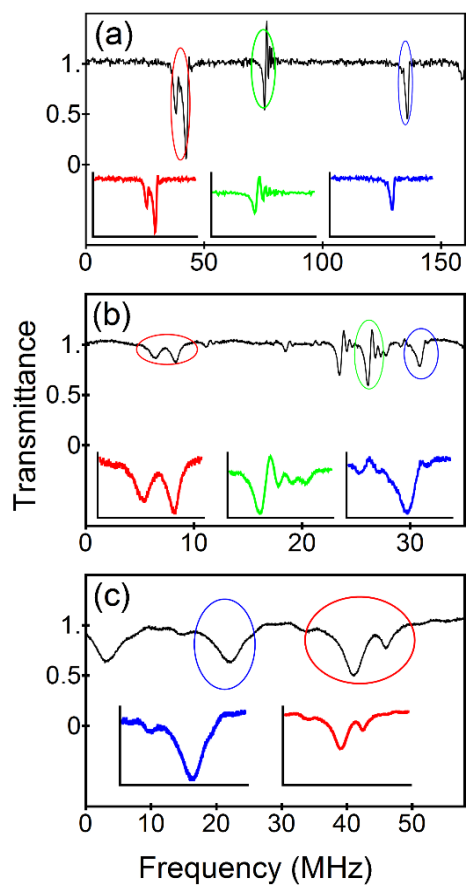

**Figure 2.** Typical transmission spectra obtained for the solid-core microbottle resonators (a) µB1, (b) µB2 and (c) µB3 as the wavelength of the tunable external cavity diode laser was scanned. Each depicting an interesting feature previously not reported for bottle resonators: Scatter-induced mode splitting (red spectrum), ringing phenomenon (green spectrum), very high-quality factor resonance (blue spectrum).

**FIGURE 3**

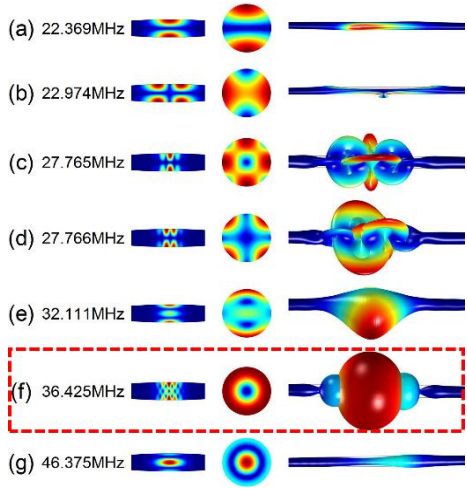

**Figure 3.** Results of numerical simulations with COMSOL Multiphysics showing seven fundamental mechanical modes that can be excited in the microbottle resonator used in our experiments. The mechanical mode shown in (f) corresponds to the mode observed in our experiment.

**FIGURE 4**

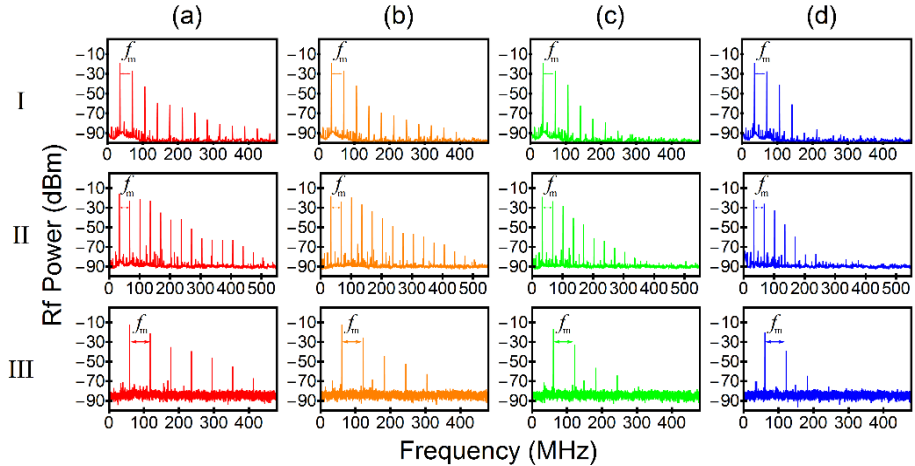

**Figure 4.** Experimentally obtained spectra of the mechanical oscillations in solid-core microbottle resonators (I) µB1, (II) µB2 and (III) µB3 using an electrical spectrum analyzer (ESA) at various values of pump power. For µB1 in row (I) the spectra were obtained at pump powers of (a) 24.1 mW, (b) 22.8 mW, (c) 20.8 mW, and (d) 20.1 mW. For µB2 in row (II) the spectra were obtained at pump power of (a) 21.2 mW, (b) 16.2 mW, (c) 12.2 mW, and (d) 8.7 mW. For µB3 shown in row (III) the spectra were obtained at pump powers of (a) 31.4 mW, (b) 19.1 mW, (c) 17.4 mW, (d) 16.8 mW. The mechanical frequencies in µB1, µB2 and µB3 were $f_m = 35.6$ MHz, $f_m = 33.7$ MHz and $f_m = 58.9$ MHz, respectively. Spectral lines separated by $f_m$ are the higher harmonics. The number of observed harmonics decreased with decreasing pump power: From 13 in (a) to 5 in (d) for µB1 (I), from 15 in (a) to 7 in (d) for µB2 (II), and from 7 in (a) to 3 in (d) for µB3 (III) respectively.

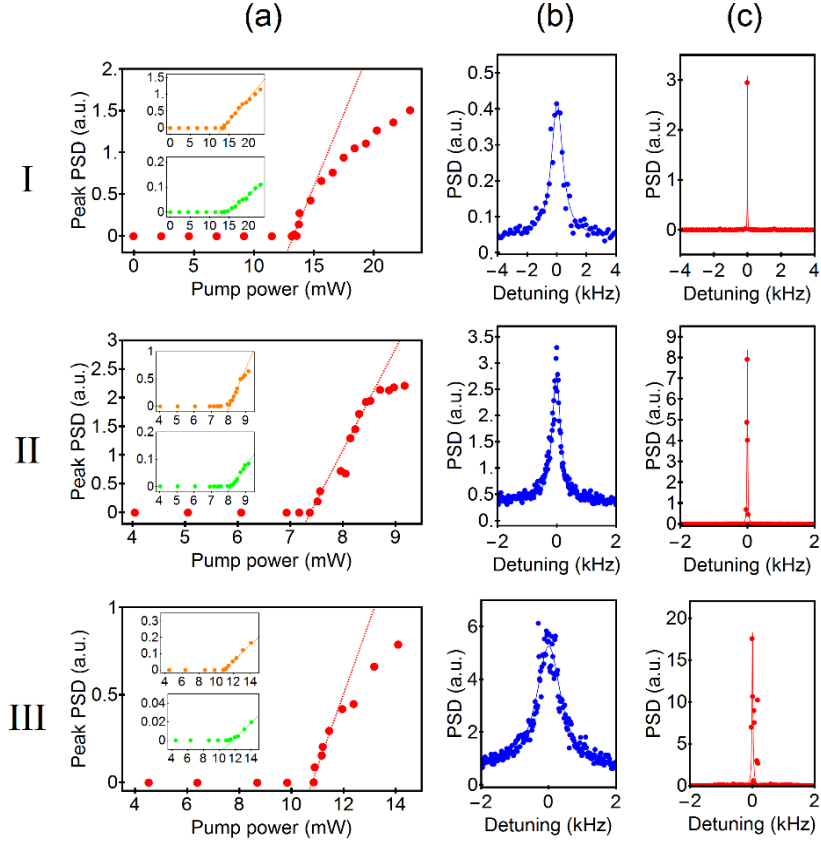

**Figure 5.** Phonon lasing in solid-core microbottle WGM resonators (I) μB1, (II) μB2 and (III) μB3. Column (a) depicts the peak of the power spectral density (PSD) versus pump power for the fundamental (red), second (inset orange) and third (inset green) harmonics. Threshold behavior and gain saturation are clearly seen. Columns (b) and (c) present, respectively, measured PSD for the fundamental mechanical modes of the microbottles below and above lasing threshold. The expected linewidth narrowing above lasing threshold is clearly seen for all resonators. The PSD shown in Ib, IIb and IIIb were obtained below lasing thresholds at pump powers of 7.8 mW, 5.4 mW and 11.2 mW, respectively, with the corresponding linewidths of 0.92 kHz, 0.33 kHz and 0.89 kHz. Above lasing threshold results shown in Ic, IIc and IIIc were obtained at pump powers of 18.4 mW, 9.7 mW and 18.2 mW, respectively. Linewidths above lasing threshold were so narrow that it could not be resolved with our electrical spectrum analyzer.

**FIGURE 6**

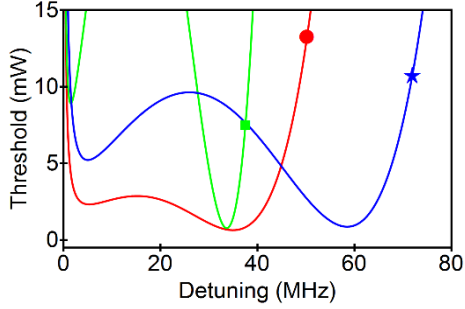

**Figure 6.** Effect of the frequency detuning $\delta$ between the pump laser and the cavity resonance on the threshold power for phonon lasing for the microbottles µB1 (red), µB2 (green) and µB3 (blue) used in the experiments. The solid lines show the threshold power of the cavity internal field calculated from Eq. (1). The markers (filled circle, square and star) on the solid curves denote the threshold powers measured in the experiment.

**FIGURE 7**

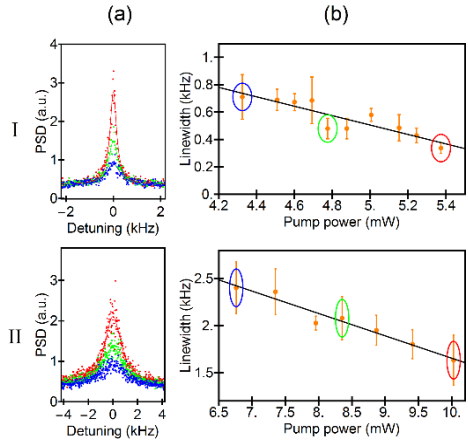

**Figure 7.** Effect of the pump power on the PSD (a) and the linewidth (b) of the mechanical modes $f_m = 33.7$ MHz in the microbottle resonator µB2 (I) and $f_m = 58.9$ MHz in the microbottle resonator µB3 (II). The PSD shown in red, green and blue colors in (Ia) and (IIa) indicate mean spectra of ten measurements and the Lorentzian fit. The linewidth versus pump power data in (Ib) and (IIb) represent the mean and the standard deviation of ten measurements performed at each pump power. The circled data points correspond to the linewidths of the spectra with the corresponding color shown in (Ia) and (IIa). The black solid line is the linear fit to the experimentally obtained data.

**FIGURE 8**

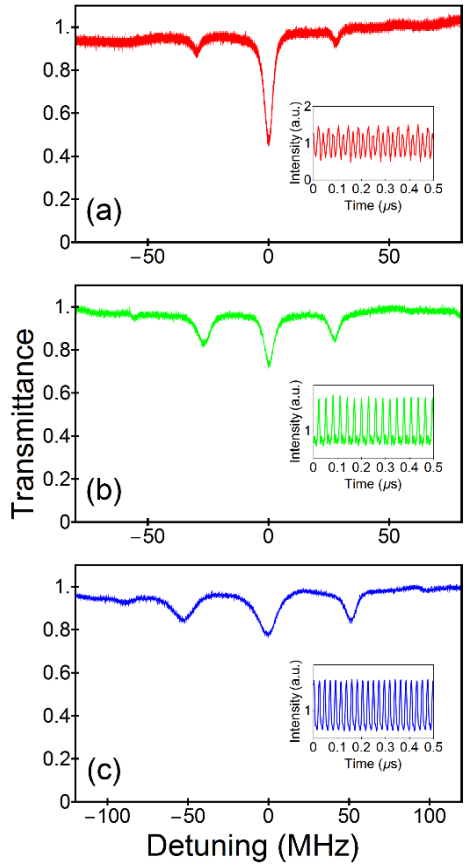

**Figure 8.** Resolved sideband spectra of the mechanical oscillation at (a) 35.6 MHz in µB1, (b) 33.7 MHz in µB2, and (c) 58.9 MHz in µB3. The insets show the time-domain signals. The first-order sidebands from the optomechanical coupling are clearly resolved in the transmission spectra.

**FIGURE 9**

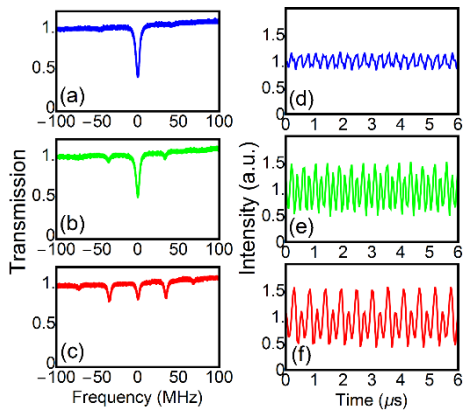

**Figure 9.** Resolved sideband spectra of the mechanical oscillation at 35.6 MHz in µB1 obtained at various values of pump power, (a) 14.0 mW, (b) 18.7 mW, and (c) 21.3 mW. (d), (e), and (f) depict the corresponding time-domain signals. At higher powers both fundamental and high-order sidebands are clearly resolved in the transmission spectra.